\documentclass[12pt,a4paper]{revtex4}
\usepackage[dvips]{graphicx}

\begin{document}

\title{Four-vertex Model and Random Tilings}
\author{N.M. Bogoliubov}

\affiliation{\ St.~Petersburg Department of V.A. Steklov
Mathematical Institute, 27, Fontanka, St.~Petersburg 191023,
Russia}

\begin{abstract}
The exactly solvable four-vertex model on a square grid with the
different boundary conditions is considered. The application of
the Algebraic Bethe Ansatz method allows to calculate the
partition function of the model. For the fixed boundary conditions
the connection of the scalar product of the state vectors with the
generating function of the column and row strict boxed plane
partitions is established. Tiling model on a periodic grid is
discussed.
\end{abstract}

\maketitle

\section{Introduction}

The connection of integrable models with the enumerative
combinatorics is widely discussed. The six-vertex model plays an
important role in these considerations. The model was widely
studied both for periodic boundary conditions and for the fixed
ones \cite {bax,vk,kz,bpz,ar,zv}. The six-vertex model for the
so-called fixed boundary conditions is related to the enumeration
of domino tilings of Aztec diamonds and of alternating sign
matrices \cite{kup,cp,fs}.

The four-vertex model is a particular case of the six-vertex model
in which two vertices are frozen out. This model was considered in
\cite{wid, park} where the connection of this model with a random
tiling models on a semi-infinite strip geometry was discussed.
Tiling models are classical discrete statistical models in which
the tiles of the different geometric shapes are packed together so
that they cover space without holes or overlaps. They are called
random only to emphasize the difference with perfect
quasi-periodic tilings.

In this paper we shall solve the model by the direct application
of the Quantum Inverse Scattering Method (QISM) \cite{fad, kbi}.
This approach will allow us to study the model on a finite lattice
with the different boundary conditions and to calculate partition
function of both vertex and corresponding tiling models. In
particular, we shall show  that the partition function of the
four-vertex model with the fixed boundary conditions is related to
the generating function of the boxed plane partitions. The boxed
plane partitions are equivalent to the three dimensional Young
diagrams placed into a box of a finite size and to lozenge tiling
of an semiregular hexagon \cite{macd, bres}. For the periodic
boundary conditions the four-vertex model is equivalent to the
lozenge tiling of the torus.

The connection of the boson type integrable models with the theory
of symmetric functions \cite{macd} and with the theory of plane
partitions \cite{ver} was considered in the papers
\cite{nmb,cel,uch,nmbt}.

The paper is organized as follows. We give the definition of the
model and discuss the QISM approach to its solution in Sec. II. In
Sec. III the model is studied on the lattice with the fixed
boundary conditions and its connection with the plane partitions
is established. The partition function of the model on the
periodic grid is calculated in Sec. IV.

\section{Four-vertex model}

A special case of the six vertex model - a four vertex model on a square
grid is described by four different arrows arrangements pointing in and out
of each vertex. A statistical weight corresponds to each type of the
vertices and there are three vertex weights $\omega _a$ , $\omega _b$ and $%
\omega _c$ (FIG. 1). There is an alternate description of the vertices in
terms of lines flowing through the vertices. Since a lattice edge can exist
in two states, line or no line, there exists a one-to-one correspondence
between the arrow configuration on the lattice and the graphs of lines on
the lattice - nests of lattice paths.
\begin{figure}[h]
\centering
\includegraphics{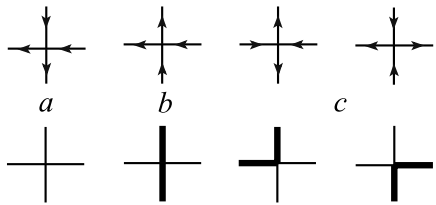}
\caption{The four allowed types of vertices in terms of arrows and lines.}
\end{figure}

For the homogeneous model the vertex weights do not depend on the
position of the vertex and the partition function is equal to
\begin{equation}
Z(\omega _a,\omega _b,\omega _c)=\sum \omega _a^{l^a}\omega
_b^{l^b}\omega _c^{l^c},  \label{partf}
\end{equation}
where the summation is extended over all allowed configurations of
the arrows on a lattice. These configurations depend on the
imposed boundary conditions which are specified by the direction
of the arrows on the boundary of the grid. The number of vertices
$(a),(b),(c)$ in each configuration is $l^a,l^b,l^c$. For the
inhomogeneous model the vertex weights are site dependent.

To apply the Quantum Inverse Scattering Method we use the spin description
of the model. With each vertical bond of the grid and horizontal bond one
associates the space $\mathcal{C}^2$, with spin up and spin down states
forming a natural basis in this space. The spin up state on the vertical
bond corresponds to the line pointing up, while the spin down state to the
line pointing down. The spin up state on the $i$-th horizontal bond $\left(
\begin{array}{c}
1 \\
0
\end{array}
\right) _i\equiv |\leftarrow \rangle _i$ corresponds to the horizontal line
pointing to the left, spin down state to the line pointing to the right $%
\left(
\begin{array}{c}
0 \\
1
\end{array}
\right) _i\equiv |\rightarrow \rangle _i$. The total space of the
vertical lines is $\mathcal{V}=(\mathcal{C}^2)^{\otimes N}$ and we
shall call it an
auxiliary space. The total space of the horizontal lines is $\mathcal{H}=(%
\mathcal{C}^2)^{\otimes N}$, and we shall call it a quantum space.
With each
vertex of the lattice one associates an operator acting in the full space $%
\mathcal{V}\otimes \mathcal{H}$. This operator is called
$L$-operator and it acts nontrivially only in a single vertical
space $\mathcal{C}^2$ and in single horizontal space
$\mathcal{C}^2$, while in all other spaces it acts as the unity
operator.

The $L$-operator of the four vertex model is equal to
\begin{equation}
L(n|u)=-\frac u2(1+\sigma ^z)e_n+ \frac {u^{-1}}{2}(1-\sigma
^z)e_n+\sigma ^{+}\sigma _n^{-}+\sigma ^{-}\sigma _n^{+},
\label{lop}
\end{equation}
where $\sigma ^{z,\pm }$- are the Pauli matrices, the projector
$e=\frac 12(\sigma ^z+1)$,  and the matrix with subindex $n$ acts
nontrivially only in the $n$-th space: $s_n=I\otimes ...\otimes
I\otimes s\otimes I\otimes ...\otimes I$.

One can represent the matrix elements of the introduced $L$-operator
\begin{equation}
L(n|u)=\left(
\begin{array}{cc}
L_{11}(n|u) & L_{12}(n|u) \\
L_{21}(n|u) & L_{22}(n|u)
\end{array}
\right) =\left(
\begin{array}{cc}
-ue_n & \sigma _n^{-} \\
\sigma _n^{+} & u^{-1}e_n
\end{array}
\right)  \label{lopp}
\end{equation}
as a dot with the attached arrows (see FIG. 2). 
\begin{figure}[h]
\centering
\includegraphics{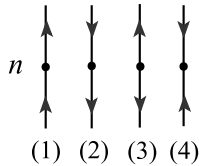}
\caption{Vertex representation of the matrix elements of the $L$-operator.}
\end{figure}
The matrix element $L_{11}(n|u)$ corresponds to a vertex $(1)$
(FIG. 2), where a dot stands for the operator $-ue_n$, this
operator acts on the local spin state and the only non-zero matrix
element of this operator is $_n\langle \leftarrow
|-ue_n|\leftarrow \rangle _n$ what gives the vertex $(b)$ (FIG. 1)
with a weight $-u$. The matrix element $L_{22}(n|u)$ corresponds
to a vertex ($2$) (FIG. 2), where a dot is the operator
$u^{-1}e_n$ and the non-zero matrix element $_n\langle \leftarrow
|u^{-1}e_n|\leftarrow \rangle _n$ is the
vertex $(a)$ (FIG. 1) with a weight $u^{-1}$. The matrix elements $%
L_{12}(n|u) $, and $L_{21}(n|u)$ correspond to the vertices ($3$) and ($4$)
 (FIG. 2) respectively, and the nonzero matrix elements $_n\langle \rightarrow |\sigma
_n^{-}|\leftarrow \rangle _n$ and $_n\langle \leftarrow |\sigma
_n^{+}|\rightarrow \rangle _n$ are the vertices $(c)$ (FIG. 1) with a weight $%
1 $.

The $L$-operator (\ref{lop}) satisfies the intertwining relation
\begin{equation}
R(u,v)\left( L(n|u)\otimes L(n|v)\right) =\left( L(n|v)\otimes L(n|u)\right)
R(u,v),  \label{llr}
\end{equation}
in which $R(u,v)$ is the $4\times 4$ matrix
\begin{equation}
R(u,v)=\left(
\begin{array}{cccc}
f(v,u) & 0 & 0 & 0 \\
0 & g(v,u) & 1 & 0 \\
0 & 0 & g(v,u) & 0 \\
0 & 0 & 0 & f(v,u)
\end{array}
\right) ,  \label{r}
\end{equation}
with
\begin{equation}
f(v,u)=\frac{u^2}{u^2-v^2},\,\,g(v,u)=\frac{uv}{u^2-v^2}.  \label{fg}
\end{equation}
The vertical monodromy matrix is the product of $L$-operators
\begin{equation}
T(u)=L(M|u)L(M-1|u)...L(0|u)=\left(
\begin{array}{cc}
A(u) & B(u) \\
C(u) & D(u)
\end{array}
\right).  \label{mm}
\end{equation}

The matrix entries of the monodromy matrix (\ref{mm}) are
expressed as  sums over all possible configurations of arrows with
different boundary conditions on a one-dimensional lattice with
$M+1$ sites (FIG. 3). Namely,
operator $B(u)=\sum_{k_M,...,k_1=1}^2L_{1k_M}(M|u)L_{k_Mk_{M-1}}(M-1|u)%
\ldots L_{k_12}(0|u)$ corresponds to the boundary conditions when
arrows on the top and bottom of the lattice are pointing outwards
(configuration (B)). Operator $C(u)$ corresponds to the boundary
conditions when arrows on the top and bottom of the lattice are
pointing inward (configuration (C)). Operators $A(u)$ and $D(u)$
correspond to the boundary conditions when arrows on the top and
bottom of the lattice are pointing up and down
respectively (configurations (A) and (D)). 
\begin{figure}[h]
\centering
\includegraphics{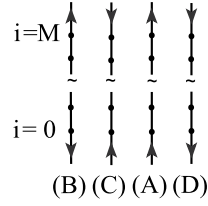}
\caption{Graphic representation of the entries of the monodromy
matrix}
\end{figure}

The commutation relations of the matrix elements of the monodromy matrix are
given by the same $R-$matrix (\ref{r})
\begin{equation}
R(u,v)\left( T(u)\otimes T(v)\right) =\left( T(v)\otimes
T(u)\right) R(u,v). \label{ttr}
\end{equation}
The most important relations are
\begin{eqnarray}
C(u)B(v) &=&g(u,v)\left\{ A(u)D(v)-A(v)D(u)\right\},  \label{cb} \\
A(u)B(v) &=&f(u,v)B(v)A(u)+g(v,u)B(u)A(v),  \nonumber \\
D(u)B(v) &=&f(v,u)B(v)D(u)+g(u,v)B(u)D(v),  \nonumber \\
\lbrack B(u),B(v)] &=&[C(u),C(v)]=0.  \nonumber
\end{eqnarray}
The transfer matrix $\tau (u)$ is the trace of the monodromy matrix in the
auxiliary space
\begin{equation}
\tau (u)=trT(u)=A(u)+D(u).  \label{trans}
\end{equation}
The relation (\ref{ttr}) means that $[\tau (u),\tau (v)]=0$ for arbitrary
values of parameters $u,v$.

The $L$-operator (\ref{lop}) satisfies the relation
\begin{equation}
e^{\zeta \sigma _n^z}L(n|u)e^{-\zeta \sigma _n^z}=e^{-\frac \zeta 2\sigma
^z}L(n|u)e^{\frac \zeta 2\sigma ^z},  \label{sls}
\end{equation}
where $\zeta $ is an arbitrary parameter. From this equation and from the
definition of the monodromy matrix (\ref{mm}) it follows that
\begin{equation}
e^{\zeta S^z}T(u)e^{-\zeta S^z}=e^{-\zeta \sigma ^z}T(u)e^{\zeta \sigma ^z},
\label{sts}
\end{equation}
where $S^z=\frac 12\sum_{i=0}^M\sigma _i^z$ is the operator of $z$ component
of the total spin. The equalities
\begin{eqnarray}
S^zB(u) &=&B(u)\left( S^z-1\right) ,  \label{bcs} \\
S^zC(u) &=&C(u)\left( S^z+1\right) ,  \nonumber
\end{eqnarray}
are the consequence of the equation (\ref{sts}). It means that the operator $%
B(u)$ decreases the total spin of the system, while $C(u)$ increases it.
From the equation (\ref{sts}) follows the commutativity
\begin{equation}  \label{taus}
[\tau(u),S^z]=0.
\end{equation}

The generating state in the space $\mathcal{H}$ is the state with all spins
up
\begin{equation}
|\Leftarrow \rangle =\otimes _{i=0}^M|\leftarrow \rangle _i=\otimes
_{i=0}^M\left(
\begin{array}{c}
1 \\
0
\end{array}
\right) _i . \label{gs}
\end{equation}
It is annihilated by the operator $C(u)$%
\begin{equation}
C(u)|\Leftarrow \rangle =0,  \label{cv}
\end{equation}
and it is an eigenvector of operators $A(u)$ and $D(u)$,
\begin{equation}
A(u)|\Leftarrow \rangle =\alpha (u)|\Leftarrow \rangle
;\,\,\,D(u)|\Leftarrow \rangle =\delta (u)|\Leftarrow \rangle ,  \label{adv}
\end{equation}
with the eigenvalues
\begin{equation}
\alpha (u)=(iu)^{M+1},\,\,\,\delta (u)=(\frac iu)^{M+1}.  \label{ev}
\end{equation}
The total spin of the generating state is $\frac 12(M+1)$:
\begin{equation}
S^z|\Leftarrow \rangle =\frac 12(M+1)|\Leftarrow \rangle .  \label{spvac}
\end{equation}

One of the main objects in the further consideration will be the vector
generated by the multiple action of operators $B(u)$ on the state $%
|\Leftarrow \rangle $
\begin{equation}
|\Psi _n(u_1,u_2,...,u_N)\rangle =\prod_{i=1}^NB(u_i)|\Leftarrow
\rangle . \label{bbb}
\end{equation}
From (\ref{bcs}) it follows that the total spin of this vector is
\begin{equation}
S^z\prod_{i=1}^NB(u_i)|\Leftarrow \rangle =\frac 12\left(
M+1-2N\right) \prod_{i=1}^NB(u_i)|\Leftarrow \rangle .
\label{spbbb}
\end{equation}
The state conjugated to (\ref{bbb}) is
\begin{equation}
\langle \Psi _n(u_1,u_2,...,u_N)|=\langle \Leftarrow
|\prod_{i=1}^NC(u_i). \label{ccc}
\end{equation}
It is easy to verify that $\langle \Leftarrow |B(u)=0$.

In the coordinate representation the state vector takes the form
\begin{equation}
|\Psi _N(u_1,u_2,...,u_N)\rangle =\prod_{k=1}^NB(u_k)|\Leftarrow
\rangle =\sum \chi _\mu (u_1,u_2,...,u_N)|m_1,...,m_n\rangle.
\label{4vstv}
\end{equation}
Here we denote the state with $N$ spins down in the sites
$m_1,...,m_N$ and with $M+1-N$ spins up in the other sites by
$|m_1,...,m_N\rangle $. The wave function $\chi _\mu
(u_1,u_2,...,u_N)$ satisfies the exclusion condition. It is non
equal to zero only if numbers $\mu =(m_1,m_2,...,m_N)$
form a strict partition $M\geq m_1>m_2>...>m_N\geq 0$ and $m_i\geq m_{i+1}+2$%
. By the direct calculations it may be shown that the wave function is of
determinantal form:
\begin{equation}
\chi _\mu (u_1,u_2,...,u_N)=(-1)^{\sum m_k}\left( u_1...u_N\right)
^{M-2(N-1)}\frac{\det \left( u_j^{-2(m_k+k-N)}\right)
}{\prod_{1\leq k<j\leq N}(u_k^{-2}-u_j^{-2})}.  \label{4vdetf}
\end{equation}
In the terms of the symmetric Schur function
\[
S_\lambda (u_1,u_2,...,u_N)=\frac{\det \left( u_j^{N-k+\lambda _k}\right) }{%
\prod_{1\leq k<j\leq N}(u_k-u_j)},
\]
it has the form
\begin{equation}
\chi _\mu (u_1,u_2,...,u_N)=(-1)^{\sum m_k}\left( u_1...u_N\right)
^{M-2(N-1)}S_{\mu -\delta }\left(
u_1^{-2},u_2^{-2},...,u_N^{-2}\right) ,\label{fvwf}
\end{equation}
where $\delta $ is a partition with the elements $\delta
_j=2(N-j),\,\,j=1,...,N$, and the elements of the partition $\lambda =$ $\mu
-\delta $ are $\lambda _j=m_j-2(N-j)$ satisfy the condition $M-2(N-1)\geq
\lambda _1\geq \lambda _2\geq ...\geq \lambda _N\geq 0$.

\section{Fixed boundary conditions}

Let us consider now a model on a $2N\times (M+1)$ square lattice
with a following boundary conditions: all arrows on the left and
right boundaries are pointing to the left while the arrows on the
top and bottom of the first $N$ columns (counting from the left)
are pointing inwards and the arrows on the top and bottom of the
last $N$ ones are pointing outwards. We shall call this condition
- the fixed boundary condition.

To enumerate all possible configuration of the vertices it is more
convenient to use the description of the model in terms of flowing
lines and to represent the allowed configurations as the nests of
lattice paths. The path is running from one of the $N$ down left
vertices to the top $N$ right ones and always moves east or north.
The paths cannot touch each other and arbitrary number of
consequent steps are allowed in vertical direction while only one
step at a time is allowed in the horizontal one. If the first $N$
columns have the numbers $(-N,\ldots ,-1)$ and the last ones
$(1,\ldots ,N) $, if the lower row has the number $0$ and the
upper one $M$, then the $m$-th path is running from the vertex
$(-N+m-1;0)$ to the vertex $(m;M)$, $1\leq m\leq N$. A typical
nest of lattice paths is represented in (FIG. 4). The length of
the path is $N+M$. The number of the $(c)$ vertices in the
admissible path is $2N$ since only one step is allowed in
horizontal direction, so the number of $(b)$ vertices is $M-N+1$.
It means that the number of $(c)$ and $(b)$ vertices in the nest
is equal  to $l^c=2N^2$ and $l^b=l^a=N(M-N+1)$ respectively.

\begin{figure}[h]
\centering
\includegraphics{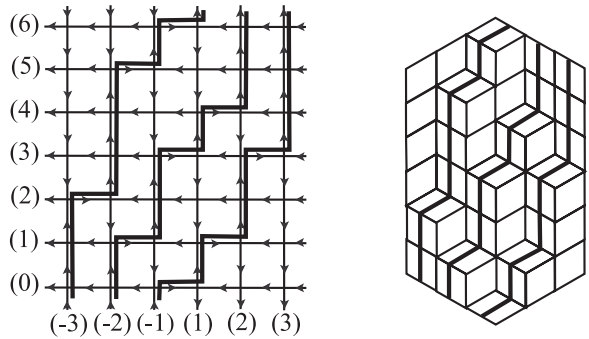}
\caption{A typical configuration of admissible lattice paths with
the fixed boundary conditions and the corresponding plane
partition with the gradient lines.}
\end{figure}

The partition function of the homogenous model with the fixed
boundary conditions is equal to
\begin{equation}
Z(\omega _a,\omega _b,\omega _c)=(\omega _a\omega _b)^{N(M-N+1)}\omega
_c^{2N^2}S(N,M),  \label{partpp}
\end{equation}
where $S(N,M)$ is the total number of the allowed nests of lattice paths.

In the inhomogeneous model the vertex weights $\omega _a$, $\omega _b$ and $%
\omega _c$ are site dependent. We shall consider a case when the
weights depend on the number of the column only. If the first $N$
columns have the numbers $(-N,\ldots ,-1)$ and the last ones
$(1,\ldots ,N)$ then the partition function is equal to
\begin{equation}
Z(\{\omega _a\},\{\omega _b\},\{\omega _c\})=\sum \prod_{k=-1}^{-N}(\omega
_a)_k^{l_k^a}(\omega _b)_k^{l_k^b}(\omega _c)_k^{l_k^c}\prod_{j=1}^N(\omega
_a)_j^{l_j^a}(\omega _b)_j^{l_j^b}(\omega _c)_j^{l_j^c}.  \label{reppf}
\end{equation}

Consider the scalar product of the state vectors (\ref{bbb}) and (\ref{ccc})
\begin{equation}
W(u_1,...,u_N;v_1,...,v_N)=\langle \Leftarrow
|C(v_1)...C(v_N)B(u_1)...B(u_N)|\Leftarrow \rangle ,  \label{scpr}
\end{equation}
where $\{u\}$ and $\{v\}$ are the sets of independent parameters.
We can represent the matrix element (\ref{scpr}) as the
two-dimensional square lattice with $2N\times (M+1)$ sites. First
$N$ vertical rows of the lattice are associated with the operators
$C(v_j)$ and the last $N$ vertical rows with operators $B(u_j)$.
The horizontal rows of the lattice are associated with the local
spin spaces, $i$-th row with the $i$-th space respectively. From
the graphical representation of the operators $B(u)$ and $C(u)$ it
follows that the matrix element (\ref{scpr}) is equal to the sum
over all allowed configurations of vertices on a square lattice
with the arrows on first $N$ vertical rows pointing inwards, on
the last $N$ ones pointing outwards; on the right and on the left
boundaries all arrows are pointing to the left (FIG. 4):
\begin{eqnarray}
W(u_1,...,u_N;v_1,...,v_N) &=&\sum
\prod_{k=-1}^{-N}(-v_{-k})^{l_k^b}(v_{-k}^{-1})^{l_k^a}%
\prod_{j=1}^N(-u_j)^{l_j^b}(u_j^{-1})^{l_j^a}  \label{wfb} \\
\ &=&(-1)^{MN}\sum
\prod_{k=-1}^{-N}v_{-k}^{l_k^b-l_k^a}\prod_{j=1}^Nu_j^{l_j^b-l_j^a}.
\nonumber
\end{eqnarray}
If we put
\begin{eqnarray}
(\omega _a)_j &=&v_{-j}^{-1},(\omega _b)_j=v_{-j};\,\,-N\leq j\leq -1,
\label{vwinm} \\
(\omega _a)_j &=&u_j^{-1},(\omega _b)_j=u_j;\,\,1\leq j\leq N,  \nonumber \\
(\omega _c)_j &=&1\,,  \nonumber
\end{eqnarray}
then
\begin{equation}
Z(\{\omega _a\},\{\omega _b\},\{\omega
_c\})=(-1)^{MN}W(u_1,...,u_N;v_1,...,v_N).  \label{pfscp}
\end{equation}

The matrix element (\ref{scpr}) for the arbitrary values of the parameters $%
u_j,v_j$ is evaluated by means of the commutation relations
(\ref{cb}) \cite{kbi, nmb} and is equal to:
\begin{equation}
W(u_1,...,u_n;v_1,...,v_n)=(-1)^{MN}\left\{ \prod_{j>k}\frac{v_jv_k}{%
v_k^2-v_j^2}\prod_{l>m}\frac{u_lu_m}{u_l^2-u_m^2}\right\} \det H,
\label{wdet}
\end{equation}
where the entries of $N\times N$ matrix $H$ are
\begin{equation}
H_{jk}=\left\{ \left( \frac{u_k}{v_j}\right) ^{M-N+2}-\left( \frac{u_k}{v_j}%
\right) ^{-M+N-2}\right\} \times \frac 1{\frac{u_k}{v_j}-\left( \frac{u_k}{%
v_j}\right) ^{-1}}.  \label{t}
\end{equation}

Formulas (\ref{wdet}), (\ref{t})give the solution of the
inhomogeneous model (\ref{wfb}) and represent its partition
function in the determinantal form.

The explicit answer for the partition function may be obtained if
we choose the weight of the vertices
\begin{equation}
v_j=q^{-\frac j2},\,\,\,u_j=q^{\frac {j-1}2},  \label{hlip}
\end{equation}
where $q=e^{-\nu }$, with $\nu >0$ is a Boltzmann weight. Under
this parametrization the partition function (\ref{reppf}) will
take a form
\begin{equation}
Z(q)=\sum q^{\sum_{k=-1}^{-N}\frac k2(l_k^a-l_k^b)+\sum_{j=1}^N\frac{j-1}%
2(l_j^a-l_j^b)},  \label{pf}
\end{equation}
where the summation is over all admissible nests of lattice paths. The
parametrization (\ref{hlip}) transforms (\ref{wdet}) into
\begin{equation}
W_q(N,M)=(-1)^{NM+\frac{N(N-1)}2}\left\{ \prod_{j>k}\left( q^{\frac{j-k}%
2}-q^{-\frac{j-k}2}\right) ^{-2}\right\} \det Q ,\label{se}
\end{equation}
with the matrix elements
\begin{equation}
{Q}_{jk}=\frac{s^{\frac{k+j-1}2}-s^{-\frac{k+j-1}2}}{q^{\frac{k+j-1%
}2}-q^{-\frac{k+j-1}2}},  \label{hme}
\end{equation}
and $s=q^{M-N+2}.$ The determinant of the matrix $Q$ was
calculated in \cite{kup} and is equal to
\begin{equation}
\det {Q}=(-1)^{\frac{N(N-1)}2}\left\{ \prod_{j>k}\left( q^{\frac{%
j-k}2}-q^{-\frac{j-k}2}\right) ^2\right\} \prod_{1\leq j,k\leq N}\frac{%
s^{\frac 12}q^{\frac{j-k}2}-s^{-\frac 12}q^{-\frac{j-k}2}}{q^{\frac{k+j-1}%
2}-q^{-\frac{k+j-1}2}}.  \label{kup}
\end{equation}
The substitution of this determinant into (\ref{se}) gives
\begin{eqnarray*}
W_q(N,M) &=&(-1)^{NM}q^{-\frac{N^2}2(M+2-2N)}\prod_{1\leq j,k\leq N}\frac{%
1-q^{M-N+2+j-k}}{1-q^{k+j-1}} \\
\ &=&(-1)^{NM}q^{-\frac{N^2}2(M+2-2N)}\prod_{1\leq j,k\leq N}\frac{%
1-q^{M+3-j-k}}{1-q^{k+j-1}}
\end{eqnarray*}
and the partition function (\ref{pf}) is given by
\begin{equation}
Z(q)=q^{-\frac{N^2}2(M+2-2N)}\prod_{1\leq j,k\leq N}\frac{1-q^{M+3-j-k}}{%
1-q^{k+j-1}}.  \label{zq}
\end{equation}

From the definition $Z(1)=S(N,M)$ and for the partition function of the
homogeneous model (\ref{partpp}) we have the following expression
\begin{equation}
Z(\omega _a,\omega _b,\omega _c)=(\omega _a\omega _b)^{N(M-N+1)}\omega
_c^{2N^2}\prod_{1\leq j,k\leq N}\frac{M-j-k+3}{j+k-1}.  \label{pfstrpp}
\end{equation}

To connect the four-vertex model with the tiling model notice that
each admissible configuration of lattice paths may be associated
with an $N\times N$ array $\pi _{ij}$. The $m-$th path may be
thought of as the $m-$th column in this array with a matrix
elements $\pi _{jm}$ equal to the number of the cells in the
subsequent columns $j$ of the lattice (starting from the right)
under the $m$-th path. The array
\begin{equation}
\pi =\left(
\begin{array}{ccc}
6 & 4 & 3 \\
5 & 3 & 1 \\
2 & 1 & 0
\end{array}
\right)  \label{ar}
\end{equation}
corresponds to the nest of paths in (FIG. 4).

An array $\pi _{ij}$ of non-negative integers that are
nonincreasing as
functions of both $i$ and $j$ $(i,j=1,2,...)$ is called a plane partition $%
\pi $ \cite{macd}, the integers $\pi _{ij}$ are the parts of the
plane partition. Each plane partition has a three dimensional
diagram which can be interpreted as  stacks of unit cubes - the
three-dimensional Young diagram, $|\pi |=\sum \pi _{ij}$ being its
volume. The height of a stack with coordinates $(i,j)$ is equal to
the part of plane partition $\pi _{ij}$. If we have $i\leq r,j\leq
s$ and $\pi _{ij}\leq t$ for all cubes of the plane partition, it
is said that the plane partition is contained in a box with side
lengths $r,s,t$. If $\pi _{ij}>\pi _{i+1j}$ , i.e. if the parts of
plane partition $\pi $ are decaying along each column, then $\pi $
is called a column strict plane partition. We shall call the
strict plane partition the partition $\pi $ that is decaying along
each column and each row ($\pi _{ij}>\pi _{i+1j}$ and $\pi
_{ij}>\pi _{ij+1}$). The element $\pi _{11}$ of the strict plane
partition $\pi _{ij}$ satisfies the condition $\pi
_{11}\geq 2r-2$ if all $i,j\leq r$. An arbitrary plane partition in a box $%
r\times r\times t$ may be transferred into a strict plane partition in a box
$r\times r\times (t+2r-2)$ by adding to an array $(\pi _{ij})$ the matrix
\[
\pi=\left(
\begin{array}{cccc}
2r-2 & 2r-3 & \cdots & r-1 \\
2r-3 & 2r-4 & \cdots & r-2 \\
\vdots & \vdots &  & \vdots \\
r-1 & r-2 & \cdots & 0
\end{array}
\right)
\]
which corresponds to a minimal strict plane partition.

A plane partition in a $r\times s\times t$ box is equivalent to a
lozenge tiling of an $(r,s,t)$-semiregular hexagon. The term
lozenge is referred to a unit rhombi with angles of $\frac \pi 3$
and $\frac{2\pi }3$ (FIG. 5).
\begin{figure}[t]
\centering
\includegraphics{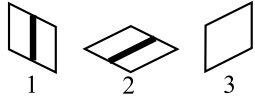}
\caption{Three types of lozenges.}
\end{figure}
The lozenge tilings are simply the projections of
three-dimensional diagrams with gradient lines. The three
dimensional diagram which corresponds to the plane partition
(\ref{ar}) is represented in (FIG. 4).

The partition function of the three dimensional Young diagrams
placed into a box is equal to
\begin{equation}
Z_{YD}(q)=\sum q^{|\pi |},  \label{yd}
\end{equation}
where $q$ is a Boltzmann weight and summation is over all plane
partitions in
a box. The volume of the three dimensional strict Young diagram in a box $%
N\times N\times M$ which corresponds to the allowed arrows (paths)
configuration is equal to \cite{nmb}:
\[
|\pi |=\frac{N^2M}2+\sum_{k=-1}^{-N}\frac k2(l_k^a-l_k^b)+\sum_{j=1}^{N}%
\frac{j-1}2(l_j^a-l_j^b).
\]
Substituting this expression into (\ref{pf}) we obtain the
partition function of the three dimensional strict Young diagrams
\begin{equation}
Z_{YD}(q)=q^{\frac{N^2M}2}Z(q)=q^{N^2(N-1)}\prod_{1\leq j,k\leq N}\frac{%
1-q^{M+3-j-k}}{1-q^{k+j-1}}.  \label{ydpf}
\end{equation}
This partition function is the generating function of the strict plane
partitions. The number of the correspondent lozenge tilings is $Z_{YD}(1)$
and is equal to the partition function of the homogeneous model (\ref{partpp}%
) with all weights equal to one:
\[
Z_{YD}(1)=Z(1,1,1)=\prod_{1\leq j,k\leq N}\frac{M-j-k+3}{j+k-1}.
\]

\section{Periodic boundary conditions}

Let us consider now the model on a lattice with the periodic
boundary conditions. It means that on a square lattice with
$(M+1)$ horizontal rows and $L$ vertical ones the boundary arrows
on any horizontal row and on any vertical one are pointing in the
same direction (see FIG. 6).
\begin{figure}[h]
\centering
\includegraphics{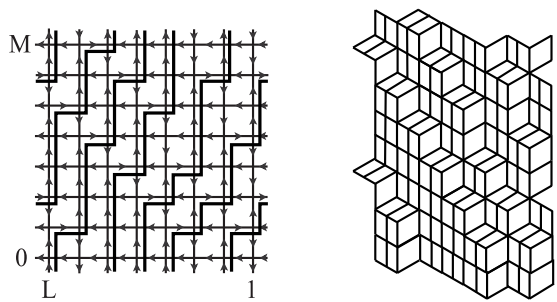}
\caption{A typical configuration of arrows and lattice paths for the
periodic boundary conditions.}
\end{figure}
Here we shall consider the case when $(M+1)$ and $L$ are even. In the
four-vertex model the number of arrows pointing to the right or to the left
is conserved in the subsequent columns of the grid. It means that the number
of vertices $c$ per column is also conserved and $l_c=2n$, where $n$ is the
number of arrows pointing to the right in a column.

The partition function (\ref{partf}) of the homogeneous model on a periodic $%
(M+1)\times L$ lattice with a fixed number of arrows pointing to
the right and the vertex weights chosen to be $\omega _a=e^{-\nu
},\,\omega _b=e^\nu $ and $\omega _c=1$ is equal to
\begin{equation}
Z_n(\mu )=\sum e^{\nu (l_b-l_a)},  \label{partfp}
\end{equation}
where the summation is carried out over all allowed configurations
of arrows on a lattice with the periodic boundary conditions or
over all nests of the admissible lattice paths (FIG. 6).

Consider the trace of the transfer matrix (\ref{trans}):
\begin{equation}
tr\tau ^L(u)=tr\left( A(u)+D(u)\right) ^L=\sum \langle
m_1,...,m_n|\left( A(u)+D(u)\right) ^L|m_1,...,m_n\rangle .
\label{trtau}
\end{equation}
Here we denote  the state with $n$ spins down in the sites
$m_1,...,m_n$ and with $M+1-n$ spins up in the other sites by
$|m_1,...,m_n\rangle$, and the sum is taken over the complete set
of states $|m_1,...,m_n\rangle $ in the quantum space
$\mathcal{H}$. The number $n$ is conserved because of the
commutativity (\ref{taus}). Since $\tau ^L(u)$ is a generating
function of different combinations of $A$ and $D$ operators, the
trace (\ref{trtau}) is the sum over all allowed configurations of
vertices $(a),(b),(c)$ with the weights $\delta (u),\alpha (u),1$
(\ref{ev}) on a periodic $(M+1)\times L $ grid
\begin{equation}
tr\tau ^L(u)=\sum (iu)^{l^b}(iu^{-1})^{l^a}.  \label{adtau}
\end{equation}
Comparing this expression with (\ref{partfp}) we obtain the following
representation for the partition function
\begin{equation}
Z_n(\nu )=e^{-i\frac \pi 2(M+1-2n)L}tr\tau ^L(-ie^\nu ),
\label{ztau}
\end{equation}
where we put $u=-ie^\nu $.

To calculate this expression one has to solve the eigenvalue
problem for the transfer matrix $\tau (u)$ which means finding
such a vector $|\Psi _n(v_1,...,v_n)\rangle $ that
\begin{eqnarray}
\tau (u)|\Psi _n(v_1,...,v_n)\rangle &=&\Theta _n(u;v_1,...,v_n)|\Psi
_n(v_1,...,v_n)\rangle ,  \label{evtm} \\
S^z|\Psi _n(v_1,...,v_n)\rangle &=&\frac 12(M+1-2n)|\Psi
_n(v_1,...,v_n)\rangle .  \nonumber
\end{eqnarray}
The vector $|\Psi _n(v_1,...,v_n)\rangle $ is an eigenvector of the transfer
matrix if parameters $v_j$ satisfy Bethe equations
\begin{eqnarray}
(v_k^2)^{M+1-n} &=&(-1)^{n-1}V^{-2},\,\,\,k=1,...,n.  \label{bth} \\
V^{-2} &=&\prod_{j=1}^nv_j^{-2}.  \nonumber
\end{eqnarray}
By putting $v_k^2=e^{ip_k}$ we can express solutions of the Bethe equations (%
\ref{bth}) in the form
\begin{equation}
p_k=\frac{2\pi I_k-P}{M+1-n},\,\,-\pi <p_k\leq \pi  \label{sol}
\end{equation}
where $I_n$ are integers or half-integers depending on $n$ being
odd or even, the total momentum $P=\sum_{j=1}^np_j$. These
solutions were classified in \cite{gaudin, abpr}.

The eigenvalue of the transfer matrix is equal to
\begin{equation}
\Theta _n(u;\{v\})=e^{i\frac \pi 2(M+1-2n)}\left\{
u^{M+1}V^{-2}+(-1)^nu^{-(M+1-2n)}\right\} \prod_{j=1}^n\frac 1{u^2-v_j^2}.
\label{teth}
\end{equation}
The partition function of the four vertex model with the periodic boundary
conditions (\ref{partfp}) is then given by the equation
\begin{equation}
Z_n(\nu )=e^{-i\frac \pi 2(M+1-2n)L}\sum_{\{v\}}\Theta
_n^L(-ie^\nu ;\{v\}), \label{finpf}
\end{equation}
where summation is taken over all different sets of solutions of
Bethe equations. The simplest case corresponds to $n=0$. The
partition function is $Z_0(\nu )=\{2\cosh (M+1)\nu \}^L$, and
$Z_0(0)=2^L$. The other limiting case is when $2n=M+1$. In this
case we have only two sets of solutions. The
first is defined by the solutions (\ref{sol}) of Bethe equations with $I_k=k-%
\frac{n+1}2$, $k=1,...,n$ and $P=0$. The other one is given by $I_k=k-\frac{%
n+3}2$, $k=1,...,n$, and $P=-\pi $. Applying the equality
\[
\prod_{k=1}^n(a-be^{i\frac{2\pi }nk})=a^n-b^n,
\]
one can find that $Z_{\frac{M+1}2}(\nu )=2$.

The vector conjugated to the Bethe eigenvector (\ref{evtm})
\[ |\Psi _n(p_1,...,p_n)\rangle
=\prod_{k=1}^nB(e^{ip_k})|\Leftarrow \rangle ,
\]
where $p_k$ are the solutions (\ref{sol}) of Bethe equations
(\ref{bth}), is equal to
\[
\langle \Psi _n(p_1,...,p_n)|=\langle \Leftarrow
|\prod_{k=1}^nB^{\dagger }(e^{ip_k}).
\]
The operators $B(e^{ip})$ and $C(e^{ip})$ are in the involution:
$B^{\dagger }(e^{ip})=(-1)^MC(e^{ip})$.

The norm of the Bethe vectors is calculated with the help of
(\ref{wdet}). When the sets $\{u\}=\{v\}$ the diagonal elements of
the matrix $H$ are equal to
\[
H_{jj}=M-n+2.
\]
When $\{u\}=\{v\}$ are the solutions of Bethe equations (\ref{bth}) the
nondiagonal entries of matrix $H$ are units: $H_{jk}=1$. It is not difficult
to calculate the determinant
\[
\det H=(M+1)(M-n+1)^{n-1},
\]
and for the norm of any Bethe eigenvector we obtain:
\[
\mathcal{N}^2(p_1,...,p_n)=\langle \Psi _n(p_1,...,p_n)|\Psi
_n(p_1,...,p_n)\rangle =(M+1)(M-n+1)^{n-1}\prod_{j\neq k}\frac{e^{iP}}{%
e^{ip_j}-e^{ip_k}},
\]
where $p_j$ are the solutions (\ref{sol}). The Bethe vectors
(\ref{evtm}) form a complete orthogonal set \cite{abpr}.

Consider the cylinder of circumference $M+1$ and the length $L$.
The
probability that the lattice paths will enter the cylinder at the cites $%
\tilde \mu =(\tilde m_1,...,\tilde m_n);\tilde m_i\geq \tilde m_{i+1}+2$ and
leave it at the cites $\mu =(m_1,...,m_n);$ $m_i\geq m_{i+1}+2$ is equal to:
\begin{eqnarray}
&&\ \ \ \ \ \langle m_1,...,m_n|\tau ^L(-ie^\nu )|\tilde m_1,...,\tilde
m_n\rangle   \label{aver} \\
\  &=&\mathcal{N}^{-2}\sum_{\{p\}}\langle m_1,...,m_n|\Psi
_n(p_1,...,p_n)\rangle \Theta _n^L(-ie^\nu ;\{e^{ip}\})\langle \Psi
_n(p_1,...,p_n)|\tilde m_1,...,\tilde m_n\rangle   \nonumber \\
\  &=&\mathcal{N}^{-2}\sum_{\{p\}}S_{\mu -\delta }\left(
e^{ip_1},e^{ip_2},...,e^{ip_n}\right) S_{\tilde \mu -\delta }\left(
e^{-ip_1},e^{-ip_2},...,e^{-ip_n}\right) \Theta _n^L(-ie^\nu ;\{e^{ip}\}),
\nonumber
\end{eqnarray}
the summation is taken over all solutions of Bethe equations, and
representation (\ref{fvwf}) has been used.

We can associate the vertical and horizontal edges of the lattice
carrying paths with lozenges. Lozenge (1) in (FIG. 5) corresponds
to a vertical line of the path, while lozenge (2) to a horizontal
one. Lattice edges without the paths correspond to lozenge (3).
The number of lozenge tilings of the torus is given by (\ref{finpf}) with $%
\nu =0$.

\section{Conclusion}

Though the considered four-vertex model is a particular case of
the six-vertex model the determinantal representation of the
partition function of the model with the fixed boundary conditions
could not be obtained as a limit because of the absence of the
correspondent answer for the six vertex case. The same is true for
the representation of the mean value (\ref{aver}) in terms of the
symmetric functions.

The quantum Hamiltonian commuting with the transfer matrix of the
four-vertex model is the Hamiltonian of the $XXZ$ Heisenberg chain
with the infinite anisotropy \cite{gaudin, abpr}. The approach
described in this paper will allow to obtain determinantal
representations for the correlation functions of the correspondent
quantum model.

\section{Acknowledgments}

The work was partially supported by the CRDF grant RUMI-2622-ST-04
and RFBR grant 07-01-00358.

\end{document}